\documentclass[english,aps, prb, twocolumn]{revtex4}
\usepackage[T1]{fontenc}
\usepackage[latin1]{inputenc}
\usepackage{amsmath}
\usepackage{graphicx}

\makeatletter

\providecommand{\tabularnewline}{\\}

\usepackage{babel}
\makeatother
\begin{document}

\title{Magnetic studies of GaN nanoceramics}

\author{A. J. Zaleski, M.Nyk, W.Strek}

\affiliation{Institute of Low Temperature and Structure Research, Polish Academy
of Sciences, Wroclaw, Poland}

\begin{abstract}
The synthesis, morphology and magnetization measurements of GaN nanoceramics
obtained under high pressure are reported. In particular the effect
of grain size on magnetic properties of GaN nanopowders and nanoceramics
was investigated. It was found that for the GaN nanoceramic characterized
by the stronger deformation of nanocrystalline grains the diamagnetic
signal changes with external magnetic field similarly to the Meissner
effect in superconductors. 
\end{abstract}
\maketitle
Gallium nitride is a wide energy gap semiconductor which is currently
massively investigated due to its application for blue laser systems.\cite{key-1}
It is well known that the GaN crystal is diamagnetic.\cite{key-2}
In last decade there were reported a number of fabrication techniques
of GaN nanocrystals.\cite{key-3,key-4,key-5} It was demonstrated
that such GaN nanopowders can be utilized as the substrate for homoepitaxy.\cite{key-6}
Recently there were developed different methods of densification of
nanopowders into large and dense ceramic bulks with grains lower than
100 nm into sizable materials retaining nanometer features. One of
the best densification techniques is the low temperature high pressure
(LTHP) method in which the compressive forces at grain boundaries
are associated with external pressure occurring at low and elevated
temperature. It results in displacement of crystalline surfaces, plastic
deformation, lattice diffusion, grain boundary diffusion and phase
metamorphosis of grain boundaries resulting in structural deformation
due to the braking of chemical bonds. In the present work we report
the magnetic properties of GaN nanoceramics obtained by LTHP method,
in particular the effect of grain sizes on the magnetic properties. 

GaN nanosized powders were synthesized by combustion method as described
earlier.\cite{key-7,key-8} The preparation procedure was divided
into two main steps - first one the hydrothermal processing using
microwaves reactor; and the second - GaN nanocrystalline powder has
been synthesized using a horizontal quartz reactor. This is a modified
method reported in our previous papers.\cite{key-9} Hydrothermal
processing using microwaves as stimulators and accelerators of the
conversion of Ga$_{2}$O$_{3}$ (Alfa Aesar 99.999\%) into Ga(NO$_{3}$)$_{3}$
was applied. The most important improvement in comparison to already
operated hydrothermal reactors was done in applying microwaves as
a heater. Waveguide is placed outside the reaction area and thus protects
the reactants from the contact with the heater elements. The Ga$_{2}$O$_{3}$
and HNO$_{3}$ were moved into the Teflon vessel, afterwards deionized
water was added. Finally prepared solution was placed in the microwave
reactor (ERTEC MV 02-02). After hydrothermal processing at 300$^{\circ}$C,
under pressure lower than 20 atm solution of Ga(NO$_{3}$)$_{3}$
was obtained. After the hydrothermal process was completed, the obtained
powder was carefully dried in an oven with gradually increasing the
temperature from 70$^{\circ}$C to 200$^{\circ}$C. Then, the powder
placed in an alumina crucible was inserted into quartz tube (24mm
ID) and was calcined at 600$^{\circ}$C for 6 h in air flow to convert
Ga(NO$_{3}$)$_{3}$ into Ga$_{2}$O$_{3}$. The crushed powders samples
were placed at room temperature into quartz tube in NH$_{3}$ flow
and after purging the material was heated to the required temperatures
850 $^{\circ}$C, 950$^{\circ}$C, and 1050$^{\circ}$C. The NH$_{3}$
vapor used for nitridation was additionally purified by passing it
over a zeolite trap. 

The fabrication of the GaN nanoceramics by means of the LTHP technique
was performed under 6 GPa at 800$^{\circ}$C for 1 min (patent application
P-378376). In this manner three samples, characterized by the different
size of the grains, were fabricated. In order to determine the structure
of the powders and ceramics the XRD diffraction measurements were
performed. Two sets of samples were prepared and studied. Overall
phase compositions of the nanopowders were determined by X-ray powder
diffraction with a Siemens D5000 diffractometer and CuK$_{\alpha1}$
radiation, $\lambda$=0.15406 nm. A nanoceramic of GaN, suitable for
crystal structure investigation, was obtained from the nanopowder
sample of GaN. The X-ray intensities data were collected on a KUMA
Diffraction KM-4 four-circle diffractometer equipped with a CCD camera,
using graphite-monochromatized MoK$_{\alpha}$ radiation ($\lambda$=
0.071073 nm). Magnetic properties of small GaN pellets with dimensions
of 1 mm thick and 4.5 mm in diameter were measured using commercial
Quantum Design SQUID magnetometer. Special care was taken to use proper
sample holder, as the measured values of magnetization of GaN were
very low. Also the effect of remnant field of superconducting magnet
was thoroughly taken care for. The samples of GaN nanoceramics were
characterized by the dimensions $\phi$4 mm and height 1.5 mm. They
were black color with average density 5.5 g/cm$^{\text{3}}$ and multipoint
BET surface area 10.5 m$^{\text{2}}$/g. 

\begin{figure}
\includegraphics[scale=0.3]{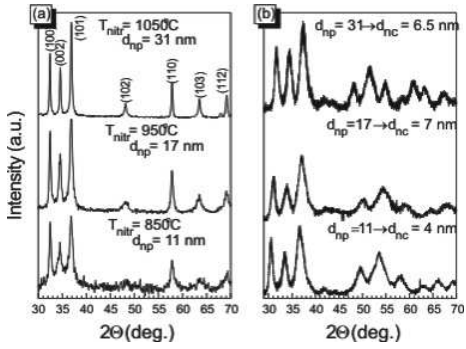}

\caption{\label{fig:1}The X-ray diffractograms of GaN powders (a) and nanoceramics
(b).}
\end{figure}

We have performed the XRD measurements of GaN nanopowders and nanoceramics.
Fig. \ref{fig:1}a shows the XRD patterns of the GaN powders obtained
at 850$^{\circ}$, 950$^{\circ}$, and 1050$^{\circ}$C, respectively.
The patterns show diffraction lines that could be ascribed to the
formation of hexagonal GaN with a wurtzite type structure (JCPDS file
No. 02-1078). By means of the Scherrer formula, there has been found
that the average size of GaN grains increases with the increase of
heating temperature. In the case of the XRD peak related to (110)
direction, the average grain sizes have been determined to be 11,
17, and 31 nm for GaN powders obtained at 850$^{\circ}$, 950$^{\circ}$,
and 1050$^{\circ}$C, respectively.\cite{key-10} It is clearly seen
that the intensities of XRD peaks increase and their broadening decrease
with the rise of heating temperature. 

Fig. \ref{fig:1}b shows XRD patterns of the GaN ceramic obtained
from the same nanopowders of GaN at 850$^{\circ}$, 950$^{\circ}$,
and 1050$^{\circ}$C, respectively. One can note that the reflexes
were much broader for the GaN nanoceramics compared to the respective
GaN powders. In case of the XRD peak related to (110) direction, the
grain sizes of nanoceramics have been determined to be 4, 7, and 6.5
nm for GaN powders obtained at 850$^{\circ}$, 950$^{\circ}$, and
1050$^{\circ}$C, respectively. We have observed some shifting of
diffraction lines. It may suggest that the structure of GaN nanocrystalline
particles after LTHP treatment was strongly deformed wurtzite structure.
Table I shows the enumerated medium-sized grains as well as the lattice
parameters for hexagonal gallium nitride. A careful analysis of the
data allows us to conclude that the average grain sizes in nanoceramics
decreased significantly due to LTHP sintering. One can conclude that
the GaN nanoceramics sintered at LTHP conditions consist of the crystalline
part, interboundary phase and amorphous phase. The volume ratio of
amorphous phase (V$_{am}$) to the volume of nanocrystalline grain
(V$_{nc}$) may be estimated assuming the spherical shape of nanograins
with d$_{np}$ for powders and d$_{nc}$ for ceramics to be V$_{am}$/
V$_{nc}$= (d$_{np}$/d$_{nc}$)$^{3}$ (see Table I). An appearance
of the amorphous phase characterizes the degree of deformation of
GaN nanoceramic responsible for creation of free carriers and their
correlations. 

\begin{table}
\begin{tabular}{|c|c|c|c|}
\hline 
GaN nanoceramic&
$d_{np}\,\left(\mathrm{nm}\right)\,\pm2$&
$d_{nc}\,\left(\mathrm{nm}\right)\,\pm2$&
$\left(d_{np}/d_{nc}\right)^{3}$\tabularnewline
\hline
\hline 
GaN (1050$^{\circ}$C)&
31&
6.5&
111\tabularnewline
\hline 
GaN (950$^{\circ}$C)&
17&
7&
14\tabularnewline
\hline 
GaN (850$^{\circ}$C)&
11&
4&
19\tabularnewline
\hline
\end{tabular}

\caption{The mean grain sizes of GaN nanopowders and nanoceramics. The symbols
are: $d_{np}$ -- mean size of grain of GaN nanopowder$^{*}$; $d_{nc}$
-- mean size of nanocrystallite of GaN nanoceramic$^{*}$ ($^{*}$calculated
using the Scherrer formula \cite{key-11}).}
\end{table}

\begin{figure}
\includegraphics[scale=0.3]{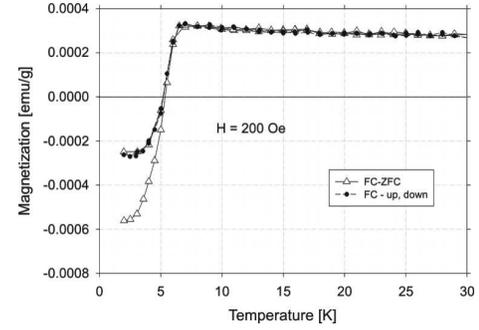}

\caption{\label{fig:2}Volume magnetization versus temperature for GaN nanoceramic
prepared from the nanopowders sintered at 1050$^{\circ}$C. (ZFC-FC
measurements with additionally sweeping temperature up and down for
FC sample).}
\end{figure}

\begin{figure}
\includegraphics[scale=0.34]{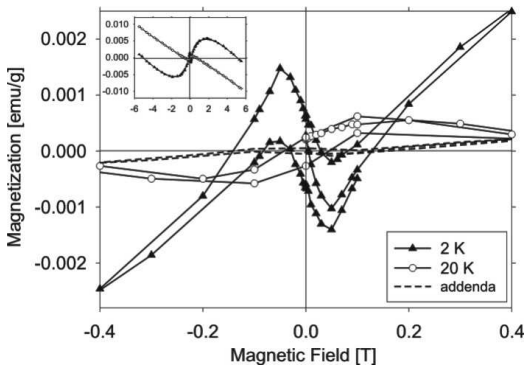}

\caption{\label{fig:3}Magnetic hysteresis for GaN nanoceramic prepared from
nanopowders obtained at 1050$^{\circ}$C ($d_{nc}$= 6.5 nm) measured
at T=2K and T=20K (signal of empty sample holder is added for comparison
-- broken line). Inset -- expanded scale of field axis.}
\end{figure}

\begin{figure}
\includegraphics[scale=0.32]{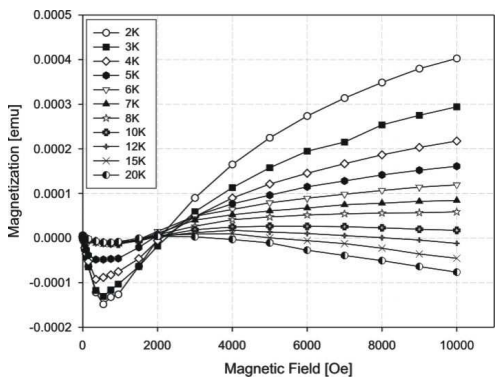}

\caption{\label{fig:4}Magnetization versus magnetic field for different temperatures
for GaN nanoceramic prepared from nanopwders sintered at 1050$^{\circ}$C.}
\end{figure}

The magnetization versus magnetic field dependence for the GaN nanoceramic
prepared from nanopowders sintered at 850$^{\circ}$C and 950$^{\circ}$C
and measured at 2K showed a paramagnetic signal, persisting up to
5T. At 20K the measured signal was practically entirely diamagnetic.
The most surprising results were obtained for magnetization versus
magnetic field measurements for GaN nanoceramics synthesized from
the nanocrystalline powders sintered at 1050$^{\circ}$C. The temperature
dependence of the field cooled zero field cooled (FC-ZFC) sample is
presented in Fig. \ref{fig:2} where for the field cooled sample the
results for measurements in increasing and decreasing temperature
are added. The behavior of GaN is typical for superconducting material
although the diamagnetic signal has similar magnitude as the paramagnetic
one. If one ascribes this diamagnetism to superconductivity, the transition
temperature T$_{c}$= 6.5 K is obtained, far above the critical temperature
of pure Gallium (T$_{c}$= 1.091 K). An additional argument for interpreting
the measured signal as superconducting behavior may be obtained from
Fig. \ref{fig:3} where the field dependence of magnetization at different
temperatures is plotted. At T = 2 K we have observed a distorted hysteresis
loop with the Meissner signal for the virgin sample. At T = 20 K this
diamagnetic signal is already absent, and as some kind of remainder
small paramagnetic signal can be observed (also hysteretic). If we
suppose that the hysteresis at T = 2 K is caused by superconductivity
appearing in the material, we can estimate the upper critical field
of the material as equal to about H$_{c2}$$\approx$0.4 T. From the
magnetization measured at different temperatures (see Fig. \ref{fig:4})
it is possible to estimate also lower magnetic field. It is equal
to about H$_{c1}$= 0.08 T at T = 2K. The ratio H$_{c2}$/H$_{c1}$=
5 suggests that our material is a weak type-II superconductor. The
superconducting-like behavior was observed only for the nanoceramics
(d$_{nc}$= 6.5 nm), i.e. for the material characterized by the largest
amorphous phase existence (V$_{am}$/V$_{nc}$$\approx$111). 

It is known that the materials subjected to high pressure demonstrate
the pressure induced metallization and superconductivity transition.\cite{key-11}
The metallization and superconductivity transitions in GaN crystal
occurring under high pressure were investigated theoretically by Nirmala
Louis et al.\cite{key-20} They are associated with enhancement of
the density of states at the Fermi level after metallization leading
to the superconductivity. The onset of superconducting transition
for GaN was predicted at 6.948 Mbar (102GPa) corresponding to the
critical temperature T$_{c}$= 0.402K. The critical temperature determined
for GaN nanoceramics was much higher (6.5K) and obtained at much lower
pressure (6 GPa). One might suppose that the high pressure applied
during the preparation of nanoceramics could lead also to similar
reorganization of the crystalline structure of the contacts between
the grains. But in such a case the superconducting behavior should
be more pronounced for nanoceramics with smaller grains, having more
numerous contacts between them. In fact from the Table I one can see
that the biggest deformation was found for nanoceramic obtained from
GaN nanopowder nitridation at 1050$^{\circ}$C. So if the observed
effects can not be ascribed to the surface, contacts or mesoscopic
effects they have probably volume character. It is tempting to explain
the superconducting behavior of GaN nanoceramics by Gallium precipitation.
But none of our XRD measurements gives any basis for stating that
free Gallium can exist in our material. Also the value of critical
temperature ascribed by us to superconducting transition of GaN nanoceramics
obtained at T = 1050$^{\circ}$C, is much higher than superconducting
transition of Gallium itself (T$_{c}$= 1.091 K). We are not aware
of any report that Gallium oxide or Gallium nitride is superconducting.
It would be also surprising if gallium precipitation was only seen
for nanoceramics obtained at 1050$^{\circ}$C. 

The structural properties of GaN nanocramics prepared by LTHP sintering
were described. It was shown that under applied pressure the nanoceramics
were strongly deformed consisting of much smaller than starting nanocrystalline
powders nanograins and amorphous phase. We have found that the GaN
nanoceramics contrary to GaN single crystals demonstrated the paramagnetic
behavior. The GaN nanoceramics characterized by the largest amorphous
phase exhibited superconductivity like behavior with the critical
temperature T$_{c}$= 5K. The nature of amorphous phase in GaN nanoceramics
and its role in creation of superconducting state needs further studies
which are in progress. 

\textbf{Acknowledgements}. The authors thank to Prof. A. Pietraszko
and Dr. M. Wolcyrz for recording XRD patterns, and Prof. W. Lojkowski
for performing hot-pressure experiment.

\end{document}